\documentclass[aps,prb,twocolumn,groupedaddress]{revtex4}

\usepackage{graphicx}
\usepackage{dcolumn}
\usepackage{bm}

\begin{document}


\title{Electronic structures of Cr$_{1-\delta}$X (X=S, Te) studied by 
Cr $2p$ soft x-ray magnetic circular dichroism}


\author{Koichiro Yaji}
\author{Akio Kimura}%
\email{akiok@hiroshima-u.ac.jp}
\author{Chiyuki Hirai}
\author{Masaki Taniguchi}
\affiliation{%
Graduate School of Science, Hiroshima University, 1-3-1 Kagamiyama, Higashi-Hiroshima 739-8526, Japan\\
}%

\author{Michie Koyama}
\affiliation{
Kure National College of Technology, Agaminami 2-2-11, Kure 737, Japan\\
}%

\author{Hitoshi Sato}
\author{Kenya Shimada}
\affiliation{
Hiroshima Synchrotron Radiation Center,Hiroshima University, 2-313 Kagamiyama, Higashi-Hiroshima 739-8526, Japan \\
}%

\author{Arata Tanaka}
\affiliation{
Department of Quantum Matter, ADSM, Hiroshima University, 1-3-1 Kagamiyama, Higashi-Hiroshima 739-8526, Japan \\
}%

\author{Takayuki Muro}
\affiliation{
Japan Synchrotron Radiation Research Institute, Mikazuki, Hyogo 679-5143, Japan \\
}%

\author{Shin Imada, Shigemasa Suga}
\affiliation{
Graduate School of Engineering Science, Osaka University, 1-3 Machikaneyama, Toyonaka, Osaka 560-8531, Japan \\
}%


\date{\today}

\begin{abstract}
Cr $2p$ core excited XAS and XMCD spectra of ferromagnetic Cr$_{1-\delta}$Te with several concentrations of 
$\delta$=0.11-0.33 and  ferrimagnetic Cr$_{5}$S$_{6}$ have been measured.
The observed XMCD lineshapes are found to very weakly depend on $\delta$ for Cr$_{1-\delta}$Te.
The experimental results are analyzed by means of a configuration-interaction cluster model calculation 
with consideration of hybridization and electron correlation effects.
The obtained values of the spin magnetic moment by the cluster model analyses are in agreement with the results of 
the band structure calculation.
The calculated result shows that the doped holes created by the Cr deficiency exist mainly in the Te $5p$
orbital of Cr$_{1-\delta}$Te, whereas the holes are likely to be in Cr $3d$ state for Cr$_{5}$S$_{6}$.
\end{abstract}


\maketitle

\section{Introduction}
Chromium chalcogenides Cr$_{1-\delta}$X (X=S, Se, Te) with metal-deficient NiAs type crystal 
structures show various magnetic and electronic properties.\cite{LB}
Among them, chromium tellurides Cr$_{1-\delta}$Te are ferromagnets with metallic conductivity
with Curie temperatures of 170-360K.

Cr$_{1-\delta}$Te with $\delta<0.1$ form the hexagonal NiAs crystal structure, 
while Cr$_{3}$Te$_{4}$($\delta$=0.25) and Cr$_{2}$Te$_{3}$($\delta$=0.33) form 
the monoclinic and the trigonal crystal structures, where Cr vacancies occupy in every
second metal layer.\cite{Ipser83}
The ordered magnetic moments evaluated from the magnetization measurements show 
much smaller values such as 2.4-2.7$\mu_{\rm B}$ for the Cr$_{1-\delta}$Te with $\delta<0.1$, 
2.35$\mu_{\rm B}$ for Cr$_{3}$Te$_{4}$($\delta$=0.25) and 2.0$\mu_{\rm B}$ for 
Cr$_{2}$Te$_{3}$($\delta$=0.33) than those expected from the ionic model.
\cite{Lotgering57, Hirone60, Ohsawa72, Grazhdankina70, Hashimoto69, Yamaguchi72, 
Andresen63, Andresen70, Hashimoto71, Kanomata00, Ohta93, Kanomata98}
According to neutron-diffraction studies, the small value of saturation magnetization 
is partly explained if we take the spin canting into account for $\delta=0.125, 0.167$ 
and $0.25$.\cite{Andresen70}
The magnetic moment induced on the Cr ion for $\delta$=0.25 is close to an integral 
number of Bohr magnetons, suggesting the existence of mixed valence Cr. \cite{Andresen70}
However for Cr$_{2}$Te$_{3}$($\delta$=0.33), the ordered magnetic moment of 
$2.65-2.70\mu_{B}$, deduced from the neutron diffraction, is much smaller than that calculated using the ionic model, 
$3\mu_{B}$, suggesting the itinerant nature of the $d$ electrons.\cite{Andresen70, Hamasaki75}

Electronic specific heat coefficients $\gamma$ are much dependent on $\delta$.
The estimated $\gamma$ for Cr$_{5}$Te$_{6}$($\delta$=0.167), Cr$_{3}$Te$_{4}$($\delta$=0.25) 
and Cr$_{2}$Te$_{3}$($\delta$=0.333) are 10, 1 and 4 mJ/atom/K$^{2}$.\cite{Gronvold64}
The $\gamma$ for Cr$_{3}$Te$_{4}$ is quite close to the predicted value 1.0$\sim$1.4 mJ/atom/K$^{2}$ 
using the calculated density of states (DOS) at the Fermi level ($E_{\rm F}$), while that for 
Cr$_{2}$Te$_{3}$ is much larger than the predicted one, $\gamma\sim0.82-0.96$ mJ/atom/K$^{2}$.\cite{Dijikstra89, Ishida}
Such large $\gamma$ values suggest that electron correlation effects are also important 
in Cr$_{5}$Te$_{6}$ and Cr$_{2}$Te$_{3}$.
The electron correlation effects in these "itinerant ferromagnets" Cr$_{1-\delta}$Te has been discussed 
with the photoemission spectra.\cite{Shimada96}
It has been pointed out that the spectral weight in $E_{\rm B}$=2-4eV observed in the photoemission 
spectra of Cr$_{0.95}$Te and Cr$_{3}$Te$_{4}$ can not be explained by the theoretical band structure 
calculation, and the intensity at $E_{\rm F}$ is found to be smaller than the theoretical DOS.
On the other hand, the spectral weight in $E_{\rm B}$=2-4eV has been reproduced with the configuration 
interaction cluster-model calculation, indicating the importance of the electron correlation effect in 
Cr$_{1-\delta}$Te.\cite{Shimada96}

In contrast to the telluride compounds, Cr$_{1-\delta}$S shows the antiferromagnetic structure and some of them 
present a metal-semiconductor transition.\cite{LB}
Among them, Cr$_{5}$S$_{6}$ shows ferri-antiferromagnetic phase transition below 150K.
In the ferrimagnetic phase above 150K, there appears a tiny net magnetic moment as a result of the 
anti-parallel spin alignment between the adjacent Cr atoms with slightly different magnetic 
moments.\cite{Larr}
Recently, in the analogy of the perovskite manganites, a collossal magnetoresistance has been found for the sulfur 
deficient Cr$_{2}$S$_{3}$, which may contain both Cr$^{3+}$($d^{3}$) and Cr$^{2+}$($d^{4}$) ions.\cite{Coey, Vaqueiro}

There has been a study of electronic structure using a band structure calculation to clarify 
the several magnetic and electronic properties depending on $\delta$ and X atom of Cr$_{1-\delta}$X (X=S, Se, Te).
\cite{Dijikstra89,Dijikstra89_2}
The calculation has successfully described the antiferromagnetic ground state for CrS, 
whereas the ferromagnetic state has not been properly described for CrTe.\cite{Dijikstra89_2}

Here we present a Cr $2p$ soft x-ray absorption (XAS) and x-ray magnetic circular dichroism (XMCD) spectra of 
ferromagnetic Cr$_{1-\delta}$Te and ferrimagnetic Cr$_{5}$S$_{6}$.
It is known that core XAS is a powerful tool to study the element specific valence electronic states of materials.
XMCD in the core XAS spectrum provides us with useful information 
on the element specific spin and orbital magnetic moments with use of the "sum rule".\cite{Thole92, Carra93, Chen95}
It is widely known that the spectral shapes of the XAS and XMCD spectra are strongly dependent 
on the electronic states or the electronic configuration of the derived atom.
Besides, these lineshapes can be remarkably affected by the intersite hybridization between the surrounding 
atoms or by the band structure of the crystal.
In other words, the spectra can be a good fingerprint of the electronic states inside the crystals.
For example, the valency (or the electron number) can be determined by the XAS and XMCD spectral lineshapes.
One can also determine several physical parameters such as the Coulomb repulsion energy, the charge transfer 
energy as well as the hybridization energy from the analyses of experimental XAS and XMCD spectral lineshapes.
In order to understand how the electronic states are related to the $\delta$ and X atom dependences of the magnetic 
and electronic properties, we have performed Cr $2p$ XAS and XMCD experiments of Cr$_{1-\delta}$Te and Cr$_{5}$S$_{6}$.

\section{Experimental}
Polycrystalline samples were synthesized from mixed powders of constituent elements.
They were sealed in evacuated silica tubes, which were heated for a week at 1000$^{\circ}C$.
After this, the samples were ground and sealed in silica tubes again and heated for 2 hours at 1450$^{\circ}$C 
and then cooled gradually to 1000$^{\circ}$C and finally quenched into water.\cite{Koyama00}
The stoichiometry and the homogeneity of Cr$_{1-\delta}$Te and Cr$_{5}$S$_{6}$ have been estimated by means of Electron 
Probe Micro-Analysis (EPMA).
X-ray diffraction studies confirmed that all of the samples were in a single phase.

Cr $2p$ core absorption spectroscopy (XAS) and X-ray magnetic circular dichroism (XMCD) spectra were measured at 
BL25SU of SPring-8 in Japan.\cite{Suga01, Suga02, Saitoh01, Saitoh02}
Circularly polarized light was supplied from a twin-helical undulator, with which almost 100\% polarization was 
obtained at the peak of the first-harmonic radiation.
After having set the two undulators to opposite helicity, helicity reversal was realized by closing one undulator 
and fully opening the other.\cite{Saitoh01}
Cr $2p$ XAS spectra were measured by means of the total photoelectron yield method by directly detecting the sample 
drain current while changing the photon energy $h\nu$.
The photon energy resolution was set to $E/\Delta E=5000$ for the Cr $2p$ core excitation regions.
The measurement was performed in the Faraday geometry with both the incident light and the magnetization perpendicular 
to the sample surface.
We used two pairs of permanent dipole magnets with holes for passing the excitation light.
The external magnetic field of $\sim1.4T$ at the sample position was alternatingly applied by setting one of the two 
dipole magnets on the optical axis by means of a moter-driven linear feedthrough.
The XMCD spectra were taken for a fixed helicity of light by reversing the applied magnetic field at each $h\nu$.
In the present paper, the XMCD spectrum is defined as $I_{+}-I_{-}$, where $I_{+}$ and $I_{-}$ represent the absorption 
spectra for the direction of magnitization (which is opposite to the direction of the majority spin) parallel and 
antiparallel to the photon helicity, respectively.\cite{Suga01}
Clean surfaces were obtained by {\it in situ} scraping of the samples with a diamond file under ultra high vacuum 
condition (3$\times$10$^{-8}$ Pa).
The cleanliness of the sample surfaces was first checked by the disappearance of a typical structure related to Cr oxides.
We could also check the degree of contamination from the magnitude of the XMCD signal, because its amplitude grew and 
finally saturated when the sample surface became clean enough.
We considered that the unscraped or contaminated surface was covered with antiferromagnetic or paramagnetic compounds 
such as Cr$_{2}$O$_{3}$, which hardly contribute to the XMCD spectrum.
It is generally known that the the total photoelectron yield reflects the absorption spectrum in the core-excitation 
region.
The temperatures during the measurement were $\sim$110K for Cr$_{1-\delta}$Te and 
200K for Cr$_{5}$S$_{6}$ (ferrimagnetic phase).

\section{CI Cluster Model Calculation with Full Multiplets}
The CI cluster model calculation has been done with a program code developed by A. Tanaka by means of the recursion method.
\cite{Recursion}
The detailed procedures are described elsewhere.\cite{Tanaka94}
Slater integrals have been calculated by Cowan's code and the calculated values are listed in Table \ref{SI}.
In the calculation, the Slater integrals are scaled down to 80\% of the listed values to take into account intra-atomic 
relaxation effect.

\section{Results and Discussion}
The XAS and XMCD spectra in the Cr $2p$ core excitation region have been measured for Cr$_{8}$Te$_{9}$ ($\delta=0.11$), 
Cr$_{5}$Te$_{6}$($\delta=0.17$), Cr$_{3}$Te$_{4}$($\delta=0.25$), Cr$_{2}$Te$_{3}$($\delta=0.33$) and 
Cr$_{5}$S$_{6}$ ($\delta=0.17$).
The Cr $2p$ XAS ($I_{+}$ and $I_{-}$) spectra with both helicities of incident radiation and the XMCD ($I_{+}-I_{-}$) 
spectra of Cr$_{5}$Te$_{6}$ are shown in Fig.1 (a) and (b).
It is found that the $2p_{3/2}$ and the $2p_{1/2}$ core absorption peaks are located at about 576 and 585 eV, and the 
broad hump is found around the photon energy of 600-605 eV.
Since the core level binding energies ($E_{\rm B}$) of the Cr $2p$ and Te $3d$ levels are quite close to each other, 
one may expect the overlapping of these absorption edges.\cite{Fuggle80}
We expect much lower Te $3d \rightarrow 5p$ absorption cross section ($<$5\%) compared to that of the Mn $2p \rightarrow 3d$
 absorption as observed, i.e., in Te $4d \rightarrow 5p$ absorption spectrum of MnTe$_{2}$.\cite{Kaznacheyev98} 

\begin{figure}
\includegraphics[width=6cm]{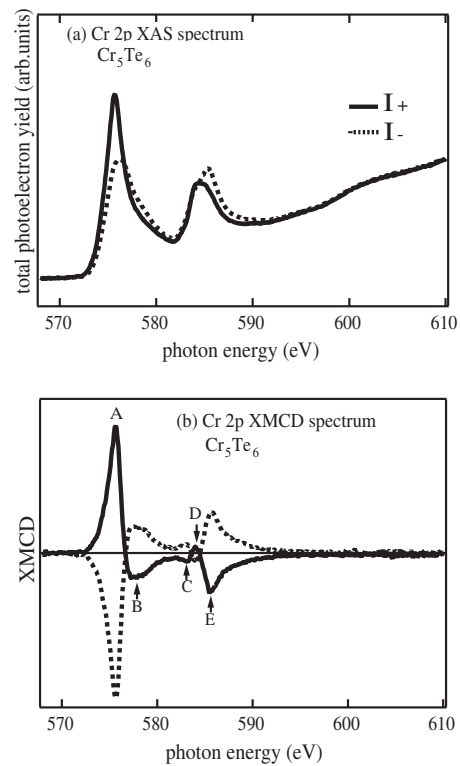}
\caption{\label{fig:epsart} (a) Cr $2p$ XAS spectra ($I_{+}$ and $I_{-}$) with different helicities of incident radiation.
(b)XMCD spectrum ($I_{+}-I_{-}$) of Cr$_{5}$Te$_{6}$ ({\it solid line}).
The XMCD spectrum taken by reversing the helicity of the incident radiation is also shown ({\it dashed line}).
}
\end{figure}

We also expect that the observed Cr $2p$ XAS fine structures are almost unaffected by the Te $3d$ XAS spectrum because 
the Te $3d$ core absorption is expected to be very broad due to the wide conduction band derived from the itinerant natre of 
the Te $5p$ electrons.
Therefore one can assume that the Te $3d$ core absorption spectrum behaves like a background and 
the observed XAS spectra in the present photon energy range mostly reflect the Cr $2p$ core absorption.
However, the broad hump at 600-605 eV can be still assigned to the Te $3d\rightarrow5p$ absorption mainly because the 
observed XMCD asymmetry is negligible in this energy region.
It is found that the intensity of $I_{+}$ is larger than that of $I_{-}$ in the $2p_{3/2}$ core absorption region, 
whereas the intensity of $I_{+}$ is smaller than that of $I_{-}$ in the $2p_{1/2}$ region.
Besides, the spectral weight of $I_{+}$ is shifted to lower energy compared to $I_{-}$ in both $2p_{3/2}$ and $2p_{1/2}$ 
regions.
This derives the complicated XMCD ($I_{+}-I_{-}$) structures as shown by the solid line in Fig.1 (b).
Here, the spectrum ({\it solid line}) shows remarkable XMCD with positive sign at the $2p_{3/2}$ core absorption edge 
($h\nu$=575.5eV) as marked with A, which is followed by the smaller asymmetry with negative sign ($h\nu$=578eV) as 
represented by B.
It is noticed that the XMCD signal does not reach zero even in the region between the spin-orbit split $2p$ components.
There is still finite and negative XMCD on the lower energy side of the $2p_{1/2}$ absorption edge ($h\nu$=583eV).
Then one finds a small positive ($h\nu$=584eV) and a large negative ($h\nu$=585.5eV) XMCD peaks with increasing $h\nu$ 
as marked with C, D and E in Fig.1 (b).
To eliminate the possible instrumental asymmetries, we have taken the spectra by reversing the helicity of the incident 
radiation.
As a result of this procedure, it is found that the observed XMCD signals with opposite helicities of the incident lights 
are quite symmetric with respect to the zero line as shown by the solid and dashed line.
This means that the observed complicated structures of the XMCD spectrum are intrinsic.

\begin{figure}
\includegraphics[width=6cm]{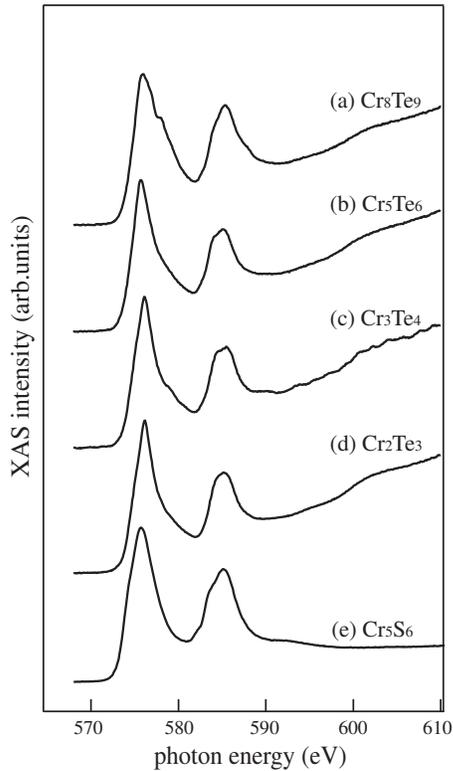}
\caption{\label{fig:epsart} Helicity averaged Cr $2p$ XAS spectra $(I_{+}+I_{-})/2$ of (a)Cr$_{8}$Te$_{9}$, (b)Cr$_{5}$Te$_{6}$, 
(c)Cr$_{3}$Te$_{4}$, (d)Cr$_{2}$Te$_{3}$ and (e)Cr$_{5}$S$_{6}$.
}
\end{figure}

In Fig.2, are shown the helicity averaged Cr $2p$ XAS spectra represented as $(I_{+}+I_{-})/2$ of 
(a)Cr$_{8}$Te$_{9}$ ($\delta$=0.11), (b)Cr$_{5}$Te$_{6}$ ($\delta$=0.17), (c)Cr$_{3}$Te$_{4}$ ($\delta$=0.25), 
(d)Cr$_{2}$Te$_{3}$ ($\delta$=0.33) and (e)Cr$_{5}$S$_{6}$ ($\delta$=0.17).
In contrast to the spectrum of Cr$_{5}$Te$_{6}$ (Fig.2 (b)), a shoulder is found on the higher $h\nu$ side of 
the $2p_{3/2}$ edge in the spectrum of Cr$_{8}$Te$_{9}$.
One also finds broad and small shoulder on the lower energy side of the $2p_{1/2}$ peaks for $\delta$=0.11.
In the cases of Cr$_{1-\delta}$Te with $\delta$=0.17, 0.25 and 0.33, the line width of the $2p_{3/2}$ edge is 
narrower with some tail extending to the higher $h\nu$ compared to that for $\delta$=0.11.
Besides, the shoulder on the lower $h\nu$ side of the $2p_{1/2}$ edge has comparable spectral weight to that of 
the $2p_{1/2}$ main peak in these spectra.
In addition, the broad humps are observed in the $h\nu$ region of 600-605eV for all of the spectra as already pointed out 
for $\delta$=0.17.
The Cr $2p$ XAS spectrum of Cr$_{5}$S$_{6}$ shows a shoulder structure at the lower $h\nu$ of the $2p_{3/2}$main peak,
which is absent in the spectra of Cr$_{1-\delta}$Te as shown in Fig.2(e).
Several structures are also found in the $2p_{1/2}$ energy region.

\begin{figure}
\includegraphics[width=6cm]{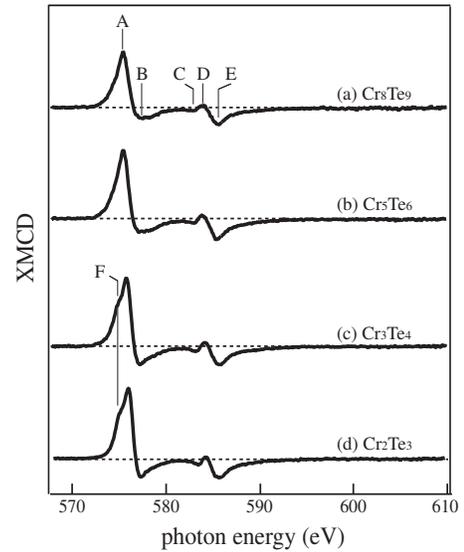}
\caption{\label{fig:epsart} Cr $2p$ XMCD $(I_{+}-I_{-})$ spectra (a)Cr$_{8}$Te$_{9}$, (b)Cr$_{5}$Te$_{6}$, (c)Cr$_{3}$Te$_{4}$ and 
(d)Cr$_{2}$Te$_{3}$.
}
\end{figure}

The XMCD spectra obtained as $I_{+}-I_{-}$ of Cr$_{1-\delta}$Te with $\delta$=0.11, 0.17, 0.25 and 0.33 are shown in Fig.3.
We find that the overall lineshape of the XMCD spectrum for $\delta$=0.11 is similar to that for $\delta$=0.17,
 in which all of the fundamental structures A-E are observed.
This is not expected because the XAS spectrum for $\delta$=0.11 shows a different feature 
from that for $\delta$=0.17 as stated above.
Therefore we speculate that the shoulder structure at the $2p_{3/2}$ edge in the XAS spectrum 
for $\delta$=0.11 (Fig.2(a)) is extrinsic and might come from the remaining part of the chromium 
oxide in the sample, which hardly contribute to the XMCD spectrum because of its non-ferromagnetic nature.
For higher $\delta$, we find that the XMCD spectrum for $\delta$=0.25 resembles to that for $\delta$=0.33.
It can be realized that the XMCD spectra for $\delta$=0.25 and 0.33 show a different feature from those 
for $\delta$=0.11 and 0.17, where the shoulder structure F is found on the low $h\nu$ side of the peak A and 
the negative structure B is sharper than those for $\delta$=0.11 and 0.17 as shown in Fig.3.

First, we have tried to evaluate the contribution of the orbital magnetic moment $m_{\rm orb}$ from the experimental XMCD 
spectra with the use of the following sum rule\cite{Thole92, Chen95},

\begin{equation}
\displaystyle m_{\rm orb}=\frac{4}{3}\cdot\frac{\int_{L_{3}+L_{2}}(I_{+}-I_{-})dh\nu}{\int_{ L_{3}+L_{2}}(I_{+}+I_{-})dh\nu}\cdot(10-n_{d})
\end{equation}
, where $n_{d}$ represents $3d$ electron number.
The estimated $m_{\rm orb}$ value is turned out to be almost negligible, suggesting the quenched Cr $3d$ orbital magnetic
 moment, in consistence with the measured $g$ value ($g\sim 2$).\cite{Grazhdankina70}
A determination of the spin magnetic moment using sum rules for lighter transition metal elements such as Cr 
is question-able because the $2p_{3/2}$ and $2p_{1/2}$ edges are not well separated and are mixed
with each other.\cite{Carra93,Teramura96}
In this case, the best way to derive the spin magnetic moment $m_{\rm spin}$ is to fit the calculated spectrum to the
experimental one as will be discussed below.

\begin{figure}
\includegraphics[width=6cm]{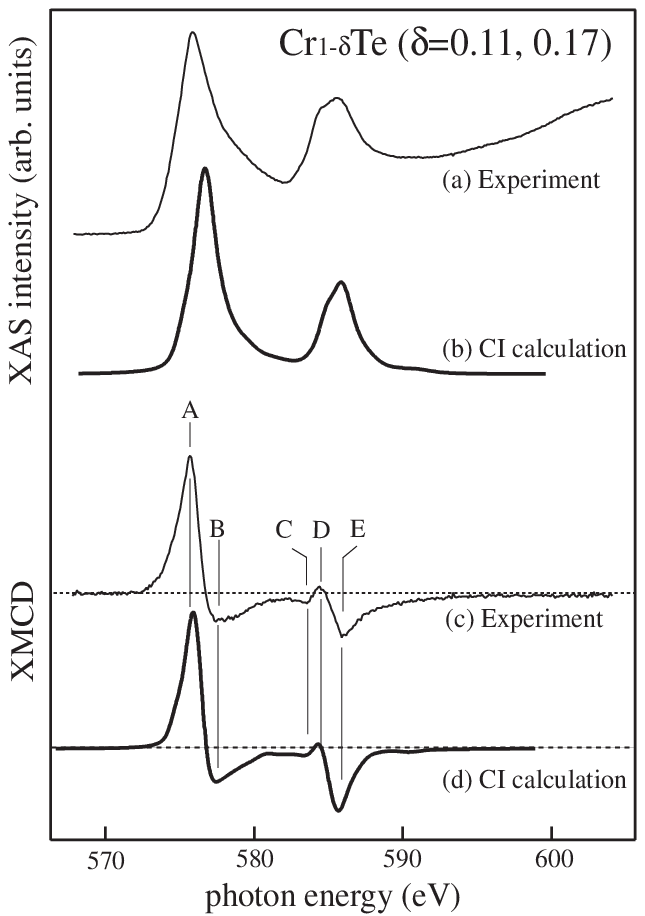}
\caption{\label{fig:epsart} (a)(c) The experimental Cr $2p$ XAS and XMCD spectra of Cr$_{5}$Te$_{6}$ ($\delta$=0.17) ({\it thin solid lines}).
(b)(d) The calculated XAS and XMCD spectra by the CI cluster model with full multiplets of [CrTe$_{6}$]$^{10-}$ 
cluster ({\it thick solid lines}).
}
\end{figure}

Next, in order to evaluate several parameters that control the physical properties of Cr$_{1-\delta}$Te, 
we have calculated the Cr $2p$ XAS and XMCD spectra by means of a configuration interaction (CI) cluster-model 
calculation with full multiplets assuming a [CrTe$_{6}$]$^{10-}$ cluster so as to reproduce the experimental 
spectra of Cr$_{1-\delta}$Te.
Here, the nominal $d$ electron numbers in Cr$_{8}$Te$_{9}$($\delta$=0.11), Cr$_{5}$Te$_{6}$($\delta$=0.17), 
Cr$_{3}$Te$_{4}$($\delta$=0.25) and Cr$_{2}$Te$_{3}$($\delta$=0.33) are $\sim$ 3.75, 3.60, 3.33 and 3.00 per Cr atom, 
respectively when we assume the Te valence to be $2-$.
\begin{table*}
\caption{\label{SI}{\it Ab initio} Hartree-Fock values of the Slater integrals and spin-orbit coupling constants (in units of eV).In the actual calculation, the Slater integrals have been scaled to 80\% of these values to take into account intra-atomic relaxation effect.}
\begin{ruledtabular}
\begin{tabular}{lc|ccccccc}
&configuration&$F^{2}(d,d)$&$F^{4}(d,d)$&$F^{2}(p,d)$&$G^{1}(p,d)$& $G^{3}(p,d)$&$\xi(3d)$&$\xi(2p)$  \\ \hline
Cr &$d^{3}$&10.777&6.755&--&--&--&0.035&--\\
&$d^{4}$&9.649&6.002&--&--&--&0.030&--\\
&$d^{5}$&8.357&5.146&--&--&--&0.025&--\\
&$d^{6}$&6.910&4.205&--&--&--&0.021&--\\
&$p^{5}d^{4}$&11.596&7.270&6.526&4.788&2.722&0.047&5.667\\
&$p^{5}d^{5}$&10.522&6.552&5.841&4.204&2.388&0.041&5.668\\
&$p^{5}d^{6}$&9.303&5.738&5.151&3.644&2.069&0.035&5.669\\
&$p^{5}d^{7}$&7.867&4.801&4.461&3.155&1.768&0.030&5.671\\
\end{tabular}
\end{ruledtabular}
\end{table*}
We have employed four charge-transfer states such as 
$d^{3}$, $d^{4}\underline{L}$, $d^{5}\underline{L}^{2}$ and $d^{6}\underline{L}^{3}$, 
where $\underline{L}$ denotes a hole in the Te $5p$ orbital.
Thus the initial state is expanded by a linear combination of these four states 
and the final state is described by a linear combination of $\underline{2p}d^{4}$, $\underline{2p}d^{5}\underline{L}$, 
$\underline{2p}d^{6}\underline{L}^{2}$ and $\underline{2p}d^{7}\underline{L}^{3}$, where $\underline{2p}$ denotes a created hole in the $2p$ core level in the 
absorption final state.
Slater integrals and spin-orbit coupling constants for $d^{3}$, $d^{4}$, $d^{5}$ and $d^{6}$ configurations in the initial state 
and for $p^{5}d^{4}$, $p^{5}d^{5}$, $p^{5}d^{6}$ and $p^{5}d^{7}$ in the XAS final states are listed in Table \ref{SI}.
To perform the CI calculation, four adjustable parameters are introduced as follows; the charge-transfer energy 
$\Delta\equiv E(d^{4}\underline{L})-E(d^{3})$, the Coulomb interaction energy $U_{dd}$ between the $3d$ electrons, 
the Coulomb attraction energy $U_{cd}$ between the $2p$ core hole and $3d$ electron, the hybridization energy $V_{\rm e_{g}} 
[=-\sqrt{3}(pd\sigma)]$ and the octahedral crystal field splitting $10Dq$.
There has been a claim that a truncated basis set using at most two configurations ($d^{n}+d^{n+1}\underline{L}$)
in the cluster model analysis would provide incorrect physical parameters especially for highly covalent materials
with small $\Delta$ value that include a highly oxidized element (such as M$^{3+}$ or M$^{4+}$) if one 
compares it with the result using a complete basis set.\cite{Mahadevan00}
Here we use four basis set that should be enough to obtain reasonable physical parameters, which would be 
confirmed from the result that the state with highest order (=$d^{6}\underline{L}^{3}$) is found to show very small 
weight ($<$7\%).

We have used the $U_{dd}\sim$2.3eV, which has been estimated from the Cr $M_{23}VV$ Auger-electron spectra and the 
self-convolution of the Cr $3d$-derived spectra.\cite{Shimada96}
The same value of $V_{\rm e_{g}}$=1.3eV [$(pd\sigma)$=-0.75eV] has been considered, where it has been evaluated by using the formula 
$\displaystyle (pd\sigma)=\eta_{pd\sigma}\frac{\hbar^{2}}{m}\cdot\frac{r_{d}^{1.5}}{d^{3.5}}$ 
with $\eta_{pd\sigma}=-2.95$ and $r_{d}$ (Cr)=0.9\AA.\cite{Harrison}
Here, $U_{cd}$ has been fixed to $U_{dd}/U_{cd}=0.83$ and the relationship $(pd\sigma)/(pd\pi)=-2.0$ has been assumed.

\begin{figure}
\includegraphics[width=6.2cm]{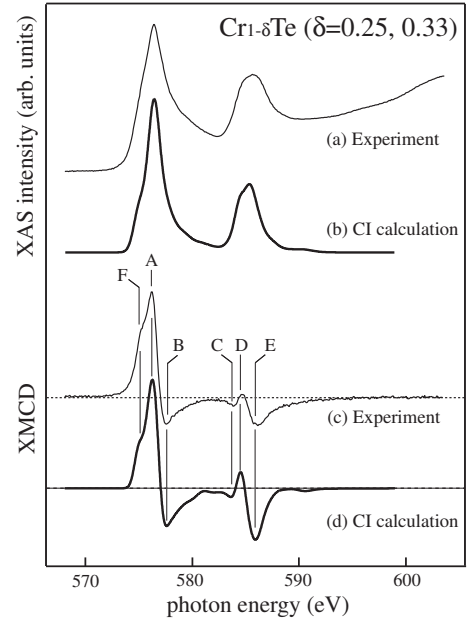}
\caption{\label{fig:epsart} (a)(c) The experimental Cr $2p$ XAS and XMCD spectra of Cr$_{2}$Te$_{3}$ ($\delta$=0.33) ({\it thin solid lines}).
(b)(d) The calculated XAS and XMCD spectra by the CI cluster model with full multiplets of [CrTe$_{6}$]$^{10-}$ 
cluster ({\it thick solid lines}).
}
\end{figure}

For Cr$_{1-\delta}$Te with $\delta$=0.11 and 0.17, we have adjusted the charge transfer energy $\Delta$ and $10Dq$ 
to fit the experimental XAS and XMCD spectra.
Fig.4 (b) and (d) show the calculated XAS and XMCD spectra with $\Delta$=-2.0eV and $10Dq$=0.7eV, 
which are broadened by the Lorenzian and Gaussian functions with the FWHM of 0.6eV for each.
It is noticed that the calculated XAS spectrum fits well with the experimental spectra for $\delta$=0.11 and 0.17 
including the shoulder structure in the $2p_{1/2}$ region.
The calculated XMCD spectrum reproduces not only the dispersive XMCD feature at the $2p_{3/2}$ edge (A and B), but 
also the structures at $2p_{1/2}$ edge including the small positive structure found in the lower $h\nu$ region (C, D and E)
 as shown in Fig.4 (d).
It is noted here that we have obtained the negative value of $\Delta$, which means that the ground 
state is not dominated by the $d^{3}$ state but by the $d^{4}\underline{L}$ and $d^{5}\underline{L}^{2}$ 
states because the energy differences $E(d^{4}\underline{L})-E(d^{3})$ and $E(d^{5}\underline{L}^{2})-E(d^{3})$ 
are expressed as $\Delta$(=-2.0eV) and $2\Delta+U_{dd}$(=-1.7eV), respectively when the hybridization ($V_{\rm e_{g}}$) 
is off.
That is, the most stable $d^{4}\underline{L}$ state is formed by the charge transfer from the ligand Te $5p$ orbitals to 
the Cr $3d$ orbitals.
The calculated result also shows that the weight of $d^{4}\underline{L}+d^{5}\underline{L}^{2}$ exceeds
80\%, which is due to the strong hybridization between Cr $3d$ and Te $5p$ orbitals.
The averaged $3d$ electron number $n_{d}$ is 4.5, which is much larger than 4 (Cr$^{2+}$) as shown in Table \ref{CI}.
The estimated $m_{\rm spin}$ and $m_{\rm orb}$ are 3.6$\mu_{\rm B}$ and -0.02$\mu_{\rm B}$, respectively.
Such a quite small $m_{\rm orb}$ stems from the dominant $d^{5}\underline{L}^{2}$ configuration in the ground state and 
is consistent with the negligible value with use of the sum rule as mentioned above.
The calculated $m_{\rm spin}$=3.6$\mu_{\rm B}$ is consistent with the values given by the band-structure calculation 
3.3$\mu_{\rm B}$\cite{Dijikstra89} but is much larger than the saturation magnetic moment $\sim$2.5$\mu_{\rm B}$.

\begin{table}
\caption{
\label{CI}Parameters obtained from the analyses of the XAS and XMCD spectra of Cr$_{8}$Te$_{9}$ and Cr$_{2}$Te$_{3}$ 
with CI cluster model calculation (in units of eV).
We assumed $U_{dd}/U_{cd}=0.83$ and $(pd\sigma)/(pd\pi)=-2$.
}

\begin{ruledtabular}
\begin{tabular}{cccccccc}
&$\Delta$&$10Dq$&$U_{dd}$&$pd\sigma$&$n_{d}$&$m_{\rm spin}$&$m_{\rm orb}$\\ \hline
Cr$_{1-\delta}$Te ($\delta$=0.11, 0.17)&-2.0&0.7&2.3&-0.75&4.5&3.6&-0.02\\
Cr$_{1-\delta}$Te ($\delta$=0.25, 0.33)&-1.5&1.0&2.3&-0.75&4.3&3.2&-0.03\\
Cr$_{5}$S$_{6}$&0.5&1.0&3.1&-0.87&3.8&3.1&-0.05\\
\end{tabular}
\end{ruledtabular}
\end{table}


We have also applied the calculation to Cr$_{1-\delta}$Te with $\delta$=0.25 and 0.33.
Figs.5 (b) and (d) show the calculated XAS and XMCD spectra with the parameters $\Delta$=-1.5eV and $10Dq$=1.0eV.
We find a good correspondence between the calculated XAS/XMCD spectra and the experimental ones.
We recognize that the calculated XMCD spectrum not only reproduces the observed fundamental structures A-E but also 
the shoulder F as shown in Figs.5 (c) and (d). 
It should be remarked here that a little bit larger $10Dq$(=1.0eV) value compared to that for $\delta$=0.11 and 0.17 
($10Dq$=0.7eV) introduced the shoulder F in the XMCD spectra for $\delta$=0.25 and 0.33.
From this result, $n_{d}$ is found to be 4.2 and the calculated $m_{\rm spin}$ and 
$m_{\rm orb}$ are estimated to be 3.2$\mu_{\rm B}$ and -0.03$\mu_{\rm B}$, respectively as listed in Table II.
We find again that the small $m_{\rm orb}$ is consistent with the XMCD result.
The present $n_{d}$ and $m_{\rm spin}$ are smaller than that for $\delta$=0.11 and 0.17.
It is noted that the value of $m_{\rm spin}$ is comparable to that obtained by the band structure calculation 
($\sim$3.3 for both $\delta$=0.25 and 0.33) \cite{Dijikstra89}, but is larger than the saturation magnetic moment
 of $\sim$2.0-2.7$\mu_{\rm B}$.


\begin{figure}
\includegraphics[width=6cm]{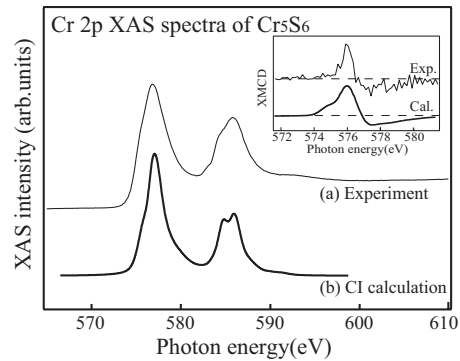}
\caption{\label{fig:epsart} (a) Helicity averaged Cr $2p$ XAS spectrum $(I_{+}+I_{-})/2$ of Cr$_{5}$S$_{6}$ ({\it thin solid lines}).
(b) The calculated XAS spectrum by the CI cluster model of  [CrS$_{6}$]$^{10-}$ cluster ({\it thick solid lines}).
The inset shows the experimental (top) and the calculated (bottom) spectra with the sama parameter set as used for 
the XAS spectrum.
}
\end{figure}

Finally the calculated results for Cr$_{5}$S$_{6}$ are shown here. 
We have evaluated the value $V_{\rm e_{g}}$=1.5eV [$(pd\sigma)$=-0.87eV] using the same formula as used for the telluride.  
It can be understood that the larger value for the sulfide is due to the reduced Cr-X atomic distance, 
where Cr-S and Cr-Te atomic distances evaluated by the X-ray powder diffraction are 2.4\AA  and 2.8\AA, respectively. 
We have adjusted the parameters so as to fit the observed spectral features as shown in Fig.6.  
The obtained parameters are $\Delta$=0.5eV, $U_{dd}$=3.1eV and $10Dq$=1.0eV.  
The value of $U_{dd}$=3.1eV is consistent with the recent photoemission and inverse photoemission experiment 
by Koyama et al., where the energy splitting between the phtoemission and inverse photoemission spectra is 
$\sim$3.0eV.\cite{Koyama03}   
It is noticed that the XMCD spectrum of Cr$_{5}$S$_{6}$ in the ferrimagnetic phase has been reproduced
with the same parameter set as shown in the inset of Fig.6, where the observed plus-minus feature is recognized
in the calculation.
The obtained positive $\Delta$ value for Cr$_{5}$S$_{6}$ is reasonable instead of the negative $\Delta$ 
for Cr$_{1-\delta}$Te because of the larger electron negativity of S than Te, which would be further confirmed 
by the reported larger value of $\Delta$=5.2eV for Cr$_{2}$O$_{3}$.\cite{Bocquet96}
In Cr$_{5}$S$_{6}$, the $d^{3}$ and $d^{4}\underline{L}$ states dominate the ground state, in contrast to the tellurides.
The evaluated average electron number $n_{d}$ is found to be 3.8, which is smaller than that of the telluride with the 
same stoichiometry.

We have obtained the similar parameter values of $\Delta$ and $10Dq$ for Cr$_{1-\delta}$Te with $\delta$=0.11-0.33,
which indicates that the local Cr $3d$ electronic states are not much affected by the Cr deficiency $\delta$.
This feature is really explained by the obtained negative value of $\Delta$, where the lowest energy state 
can be described as $d^{4}\underline{L}$.
Thus we interpret that the doped holes created by the Cr deficiency do not stay in the Cr $3d$ states but exist 
in the Te $5p$ orbitals for Cr$_{1-\delta}$Te.
This is consistent with the calculated band structure of Cr$_{1-\delta}$Te, where the hole pocket derived from the 
Te $5p$ state appears around $\Gamma$ point in the Brillouin zone when the Cr vacancy is introduced in the material.
\cite{Dijikstra89}
In contrast, the positive $\Delta$ value for Cr$_{5}$S$_{6}$ leads to the lowest energy state of $d^{3}$,
which means the doped holes are likely to be in the Cr $3d$ orbitals.
This interpretation is consistent with the ferrimagnetic resonance experiment on Cr$_{5}$S$_{6}$, where 
the electron hopping between the Cr$^{2+}$ and Cr$^{3+}$ ions is suggested.\cite{Konno88}

\section{Conclusion}
We have observed the Cr $2p$ XAS and XMCD spectra of Cr$_{8}$Te$_{9}$ ($\delta$=0.11), Cr$_{5}$Te$_{6}$ ($\delta$=0.17),
 Cr$_{3}$Te$_{4}$ ($\delta$=0.25), Cr$_{2}$Te$_{3}$ ($\delta$=0.33) and Cr$_{5}$S$_{6}$.
The observed changes with the Cr vacancy in the experimental XMCD spectra of Cr$_{1-\delta}$Te are found to be
small.
The experimental XAS and XMCD spectra of Cr$_{1-\delta}$Te have been compared with the result of 
the CI cluster-model calculation.
We have obtained the best-fit parameter values, $\Delta$=-2.0eV, $10Dq$=0.7eV for $\delta$=0.11 and 0.17 and 
$\Delta$=-1.5eV and $10Dq$=1.0eV for $\delta$=0.25 and 0.33 with $U_{dd}$=2.3eV and 
$V_{\rm e_{g}}$=1.3eV fixed for all of Cr$_{1-\delta}$Te studied.
Both the experiment and the calculation give a quenched Cr $3d$ orbital magnetic moment for Cr$_{1-\delta}$Te,
in agreement with the observed $g$ value.
The obtained values of the spin magnetic moment by the cluster model analyses are consistent with the results of 
the band structure calculation.  
In contrast to these negative $\Delta$ values for the tellurides, we have found the positive $\Delta$ (=0.5eV)
for Cr$_{5}$S$_{6}$.
It is concluded from these results that the doped holes created by the Cr deficiency exist mainly in the Te $5p$
orbitals for Cr$_{1-\delta}$Te, whereas the holes are likely to be in Cr $3d$ state for Cr$_{5}$S$_{6}$.
This interpretation for Cr$_{1-\delta}$Te is consistent with the unchanged feature of the XMCD spectrum 
with the Cr defect concentration $\delta$ and the band structure calculation.



\begin{acknowledgments}
The authors would like to thank Dr. Y. Saitoh of the Japan Atomic Energy Research Institute for a fine adjustment 
of the monochromator and Professor A. Fujimori of the University of Tokyo and Professor T. Kanomata of 
Tohoku Gakuin University for their fruitful discussion.
This work was done under the approval of the SPring-8 Advisory Committee (Proposal No. 2000B0439-NS -np).
This work was supported by the Ministry of Education, Science, Sports and Culture.
\end{acknowledgments}


%

\end{document}